\documentclass[11pt, letterpaper]{article}
\usepackage[utf8]{inputenc}
\usepackage[margin=1.5cm]{geometry}
\usepackage{graphicx}
\usepackage{titlesec}
\usepackage{array}
\usepackage{tabu}
\usepackage{longtable}
\usepackage{booktabs}

\newcolumntype{C}[1]{>{\centering\arraybackslash}p{#1}}
\newcolumntype{L}[1]{>{\raggedright\arraybackslash}p{#1}}
\usepackage{enumitem}
\usepackage{amssymb}
\usepackage{amsmath}
\usepackage{xcolor}

\newlist{selectlist}{itemize}{2}
\setlist[selectlist]{label=$\square$,leftmargin=*,noitemsep,topsep=0pt}

\usepackage{lmodern}

\usepackage{hyperref}
\hypersetup{
    colorlinks=true,
    linkcolor=blue,
    filecolor=magenta,
    urlcolor=blue,
}

\titleformat{\section}[block]{\hspace{1em}\bfseries}{\thesection.}{0.5em}{}
\titleformat{\subsection}[block]{\hspace{1em}}{\thesubsection}{0.5em}{}

\begin{document}

\setlength{\parindent}{0pt}
\setlength{\parskip}{10pt}

\begin{flushleft}

\textbf{Article title}\\
A Low-Cost MR-Compatible Pneumatic Motion Phantom for Magnetic Resonance Imaging

\textbf{Authors}\\
Jonathan B. Martin\textsuperscript{1,*}, Millie A. Barron\textsuperscript{2}, Charlotte R. Sappo\textsuperscript{1,2}

\textbf{Affiliations}\\
\textsuperscript{1} Vanderbilt University Institute of Imaging Science, Vanderbilt University Medical Center, 1161 21st Avenue South, Nashville, TN 37232, USA\\
\textsuperscript{2} Vanderbilt University, 2201 West End Avenue, Nashville, TN 37235, USA\\
\textsuperscript{*} Author to whom any correspondence should be addressed.

\textbf{Corresponding author's email address and Twitter handle}\\
jonathan.bach.martin@vumc.org\\
twitter: @jonbachmartin

\textbf{Abstract}\\
We describe the detailed design, construction, and testing of a simple, low-cost, open-source, exclusively pneumatic phantom for gated magnetic resonance imaging (MRI) studies of cyclic motion. The overall phantom cost is approximately \$200. It is designed to use widely-available materials and an easy-to-program, adaptable micro-controller. Image contrast is generated with air pumped through mineral oil. We demonstrate the utility of the phantom for simulating simple organ motion when developing and investigating image acquisition and reconstruction methods. Phantom motion was independently characterized by video analysis and found to be highly repeatable, with a cycle-to-cycle period variation of 0.6\% and a stroke volume variation of 2.3\%. The phantom does not generate electromagnetic interference, making it a suitable option for unshielded ultra-low field (ULF) MRI systems. To our knowledge, this work reports the first motion phantom to be imaged on an ULF MRI system.


\newpage
\textbf{Specifications table}
\vskip 0.2cm
\tabulinesep=1ex
\begin{tabu} to \linewidth {|X[l]|X[3,l]|}
\hline \textbf{Hardware name} & Pneumatic magnetic resonance imaging motion phantom
  \\
\hline \textbf{Subject area} &
  \begin{itemize}[noitemsep, topsep=0pt, leftmargin=*]
  \item Engineering and material science
  \item Medical (e.g. pharmaceutical science)
  \end{itemize}
  \\
\hline \textbf{Hardware type} &
  \begin{itemize}[noitemsep, topsep=0pt, leftmargin=*]
  \item Imaging tools
  \end{itemize}
  \\
\hline \textbf{Closest commercial analog} &
  MRI Compatible Multi-Modality Heart Motion Phantom (Shelley Medical Imaging Technologies, London, ON, Canada)
  \\
\hline \textbf{Open source license} &
  CC BY 4.0
  \\
\hline \textbf{Cost of hardware} &
  \$219.25 USD (\$187.13 USD excluding the modifiable mineral oil filling liquid)
  \\
\hline \textbf{Source file repository} &
  \href{https://doi.org/10.5281/zenodo.17653758}{https://doi.org/10.5281/zenodo.17653758}
  \\
\hline
\end{tabu}
\end{flushleft}

\newpage
\section{Hardware in context}

Magnetic resonance imaging (MRI) provides anatomical and functional information about organs throughout the body, making it a mainstay of clinical practice and biomedical research. MRI has significant strengths that complement those of other imaging modalities. Unlike computed tomography (CT), MRI is free of ionizing radiation and has superior soft tissue contrast (useful for applications such as cardiac and lung tumor imaging \cite{gatti_cardiovascular_2021}) and greater ability to characterize hemodynamics (useful for applications such as monitoring repair of congenital heart disease \cite{pushparajah_cardiovascular_2019}). Unlike ultrasound imaging, MRI is not constrained by the available acoustic windows that limit image geometry and field of view.

However, magnetic resonance imaging of the cardiopulmonary system is challenging because of motion, requiring specialized methods of image acquisition such as the use of gating. The heart and lungs undergo complex, non-uniform but cyclic motions in three dimensions, and the relatively slow nature of MR signal measurements makes imaging of such moving structures challenging and prone to motion artifacts. Motion artifacts from moving organs can manifest as blurring, signal intensity variations, or ``ghosting'' in images, which each have a deleterious effect on image quality. Gating techniques (either external or self-gating), rapid imaging sequences, and advanced image reconstruction algorithms have been developed and implemented to overcome motion effects to produce diagnostic-quality images \cite{zaitsev_motion_2015}.

Test objects and imaging phantoms are widely used in radiological practice for various purposes \cite{altermatt_design_2019}, including providing useful surrogate objects to test and evaluate new imaging methods or hardware prior to their application \textit{in vivo}. Such objects must replicate some realistic features of human physiology and anatomy to provide an appropriate testbed. Motion-mimicking phantoms are particularly challenging because a realistic cyclic motion and modulation of intrinsic signal contrast must be produced.

Producing these effects in the MRI environment limits the design and choice of materials because the use of many metallic materials or actuators in the phantom that might be used to generate motion is precluded. Ferromagnetic or highly paramagnetic components in particular cannot be used within some limits within the fringe field. Electromagnetic interference (EMI) that might be detected by the MRI receiver is also a concern and constrains design. For example, stepper motors, such as those used in Ref.\ \cite{weidner_abdominal_2022} produce sporadic EMI that is amplified through the receive chain \cite{srinivas_external_2022}. This can be especially problematic when used in conjunction with an MRI system that is unshielded, such as emerging low-field scanners \cite{martin_bloch-siegert_2021, wald_low-cost_2019}. The materials of the phantom itself must also be chosen to be visible on MR and to have realistic MR relaxation parameters, ideally with contrast between tissue types.

Several phantoms meant to simulate the motion of the human heart and lungs have been produced for use with MRI over the years \cite{wheatley_novel_2022, korner_modular_2024}. A number of highly realistic commercial systems are available, for example from Shelley Medical Imaging Technologies (demonstrated in Ref.\ \cite{huang_rapid_2009}) or GTSimulators. These systems are anatomically accurate and can replicate a wide range of physiological motion and flow conditions. However, these commercial phantoms are premium products costing tens of thousands of dollars and usually have limited potential for customization. A few low-cost and open-source phantoms exist, but each has limitations. A periodic motion phantom based on the K'NEX construction system (K'NEX Industries, Inc., Hatfield, PA) with hardware schematics is available \cite{huber_low-cost_2000}; however, this system produces only rigid rotational motion, which cannot be used to model volumetric heart or lung changes. Purely pneumatic cardiac phantoms have been created \cite{evans_cardiac_1988} that rely on a motor that is relatively expensive but not suitable for use with unshielded installations. Lung phantoms operating under similar pneumatic principles have recently been developed, but at unspecified or higher cost \cite{kim_technical_2022, wheatley_novel_2022}. We have identified only one fully open-source cyclic motion phantom system including a comprehensive, detailed list of construction materials and complete assembly instructions \cite{gnanago_modular_2025}. While this modular system is an exemplary demonstration of reproducible hardware design and incorporates both cardiac and respiratory motion, we aimed to further improve simplicity and reduce cost. Unlike this previous phantom, we use a simple controller designed for educational use and programmed with the Arduino C++ dialect, and rely exclusively on pneumatic tubing and an inflatable balloon to produce motion, without the need for piezo motors or customized gearing.

Here we present an open-source, low-cost, pneumatic motion phantom suitable for gated acquisition of MR images. This phantom is particularly suited for and motivated by the need to prototype sequences on low-cost ultra-low field (ULF) MRI systems, often defined as $\leq$ 0.1 T \cite{gagliardo_ultra-low-field_2025, zhao_whole-body_2024}. The phantom integrates simple pneumatic hardware, a user-friendly and flexible embedded controller, and external triggering capabilities. Although the structure and motion of the proposed phantom are less anatomically and physiologically realistic than some previously developed or commercially available phantoms, our system features a much simpler construction process than prior phantoms, at an extremely low cost (approximately \$200 in 2026 USD), making this the most affordable MRI motion phantom we have identified with disclosed costs to date.

\section{Hardware description}

The phantom is a pneumatically driven system meant to approximate the cyclic expansion and contraction motion seen in the heart. The core components of the system are 1) a microcontroller and motor control board, 2) a pneumatic system for actuating the phantom, and 3) the phantom itself, including the moving components and 3D printed housing.

\subsection{Control system}
The phantom is controlled using the Adafruit Circuit Playground Express (CPX) (Adafruit, Brooklyn, NY, USA), a flexible educational microcontroller platform for prototyping of embedded systems. The CPX board is centered around an ATSAMD21 microcontroller and has a wide array of added components, including NeoPixel LEDs, speakers, I2C, UART, analog and digital in/out, PWM, and a variety of sensor inputs (light, temperature, audio, motion, touch). The CPX facilitates several different methods of programming the board, including a graphical programming interface, Javascript programming, a Python interpreter running on the board, or the Arduino IDE. For all developments discussed here, the Arduino IDE was used, with programs written in the Arduino C++ dialect. This provides a significant cost advantage over other dynamic motion phantoms controlled using a MATLAB interface \cite{gnanago_modular_2025}. This choice avoids the cost of a MATLAB license and provides additional savings by avoiding dependency on the small number of MATLAB-supported hardware vendors, e.g. Ref.\ \cite{gnanago_modular_2025} relies on a relatively expensive MATLAB-controlled data acquisition card (approx. \$350) as just one component of the control system. An add-on board, the Adafruit CRICKIT, is used to power the pneumatic system. The CPX is directly mounted on the CRICKIT using stand-off bolts and extends the CPX to provide control of motors, servos, and solenoids. A single 5V DC power supply is used to power the CRICKIT, CPX and all controlled motors. A ring of NeoPixel programmable LEDs are included on the CPX, and changing the color of these LEDs is performed during the simulated cardiac cycle as a visual cue to the user of the phantom's motion (which is otherwise often obscured inside the MRI system in most cases).

\subsection{Pneumatic system}
\begin{figure}[htbp]
    \centering
    \includegraphics[width=1\linewidth]{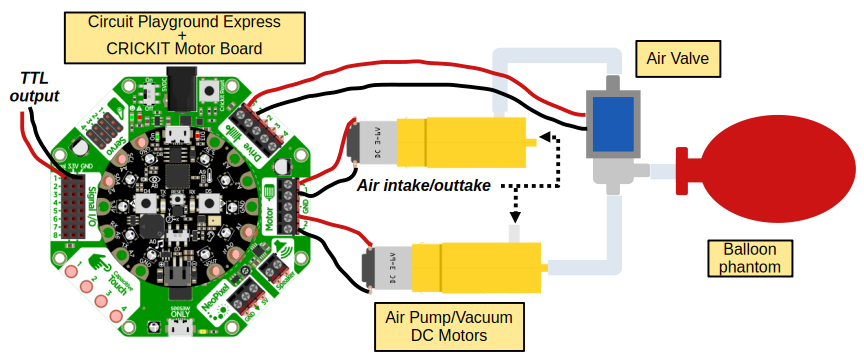}
    \caption{Schematic of the pneumatic control system. All actuation is controlled by an Adafruit Circuit Playground Express with a CRICKIT motor board.}
    \label{fig:schematic}
\end{figure}

Figure \ref{fig:schematic} shows a diagram of the pneumatics used to actuate the phantom and their interface with the CPX/CRICKIT. Two DC air pumps are used to inflate and deflate the phantom. Each air pump is capable of pumping air only in one direction (reversing motor polarity does not change the direction of flow), necessitating two pumps. These are powered by the CRICKIT's 1 A brushed DC motor control, and pumping rates are controlled via PWM. The intake and outtake pumps are connected to a three-port air release valve. The middle port is a common connection, which connects to the balloon `ventricle' phantom. Powering the valve with the CRICKIT's 500 mA drive output switches which of the two end ports are connected to the common middle port. Synchronizing the action of the valve with that of the air pumps can allow the balloon to inflate, deflate, or remain in stasis. The CRICKIT's digital outputs are used to generate a TTL signal for triggering the MR system. This TTL signal may be used for either prospectively or retrospectively gated imaging.

\subsection{Phantom housing}
\begin{figure}[htbp]
    \centering
    \includegraphics[width=1\linewidth]{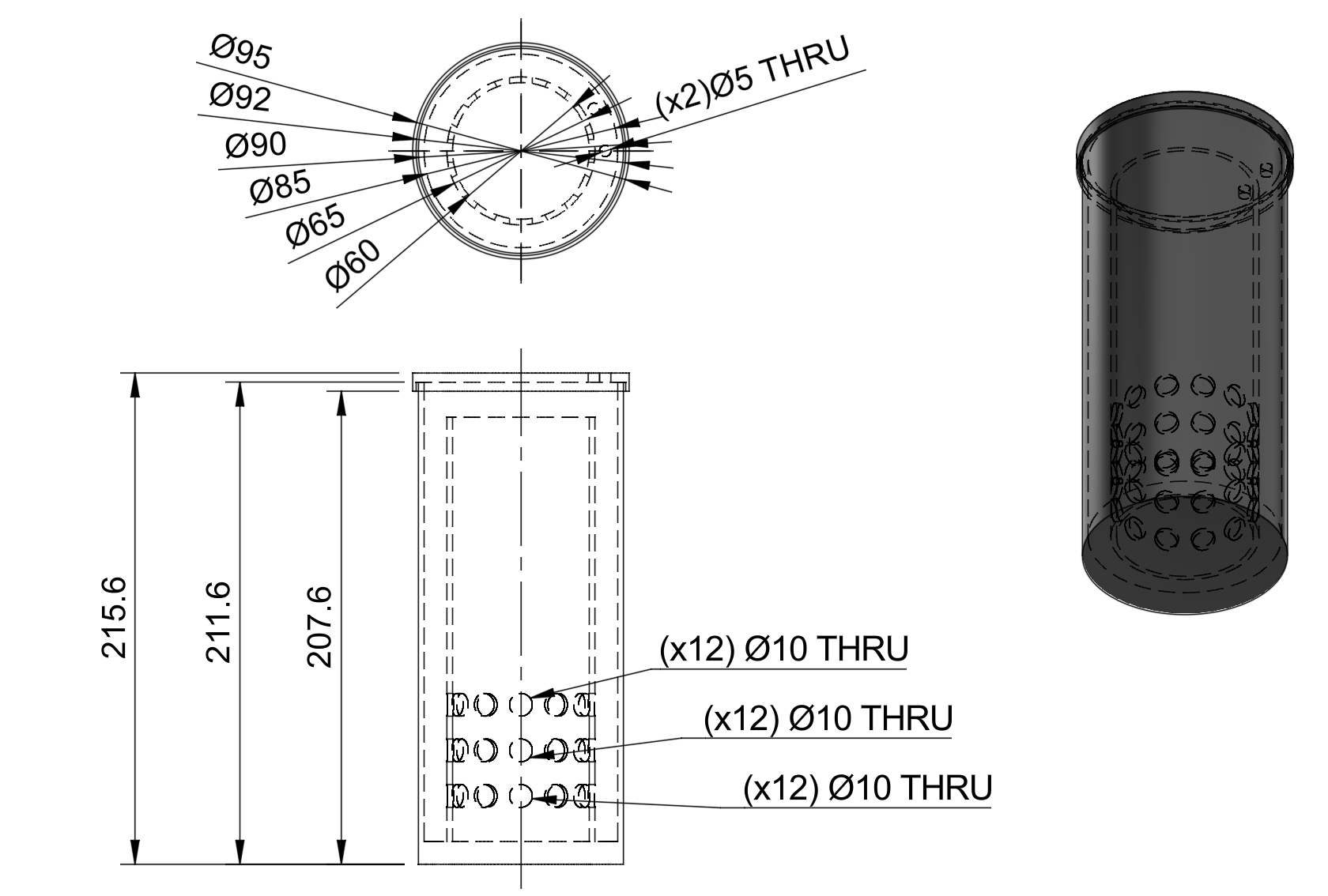}
    \caption{Phantom housing design (design file \texttt{balloon\_\allowbreak phantom\_\allowbreak V1.2}). Units are in [mm]. The housing consists of two concentric tubes, sharing a common base. Holes in the bottom section of the inner tube allow for insertion of pneumatic line and movement of filling fluid. A lid attaches to the top of the phantom, and has holes for the pneumatic line and airflow in and out of the container.}
    \label{fig:CAD}
\end{figure}

\begin{figure}[htbp]
    \centering
    \includegraphics[width=0.65\linewidth]{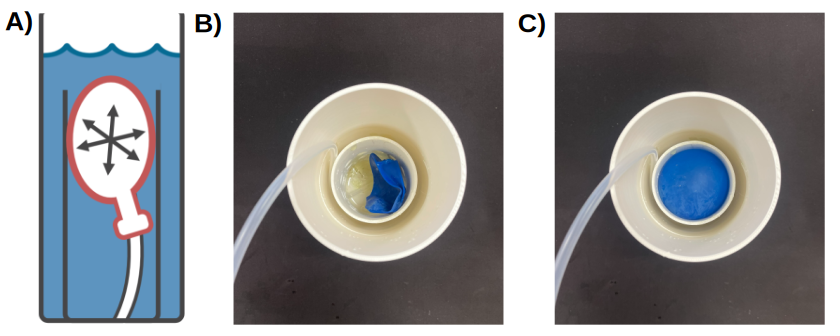}
    \caption{Phantom functionality. (A) A conceptual drawing of the phantom. (B) A top-down view of the phantom with the balloon not inflated. (C) A top-down view of the phantom with the balloon inflated to the walls of the inner compartment for validation.}
    \label{fig:function}
\end{figure}

Figure \ref{fig:CAD} shows the design of the 3D printed housing for the phantom. Image structure and contrast is produced by an air-filled balloon suspended in mineral oil, following prior ventricle phantoms that use air to generate ventricle/myocardium contrast \cite{timinger_mr_2006}. The 3D printed phantom housing consists of two concentric tubes sharing a circular base. A lid with ports for pneumatic lines and air flow is included in the CAD. The outer tube defines the extent of the phantom, while the inner tube serves to create a reproducible motion by constraining the expansion of the balloon. By constraining the outward expansion of the balloon, this inner tube also allows for verification of measured area of the balloon, by providing a known cross sectional area at the maximum inflation. Figure \ref{fig:function} demonstrates the function and motion of the phantom and shows how the balloon and the inner tube interface. The inner 3D printed tube has holes in the lower part of its curved surface to allow mineral oil to freely move back and forth between the interior and exterior tubes. Our exclusive reliance on pneumatics to drive the motion of the phantom eliminates the need for EMI mitigation while using this phantom.

\subsection{Control system housing}
\begin{figure}[htbp]
    \centering
    \includegraphics[width=0.5\linewidth]{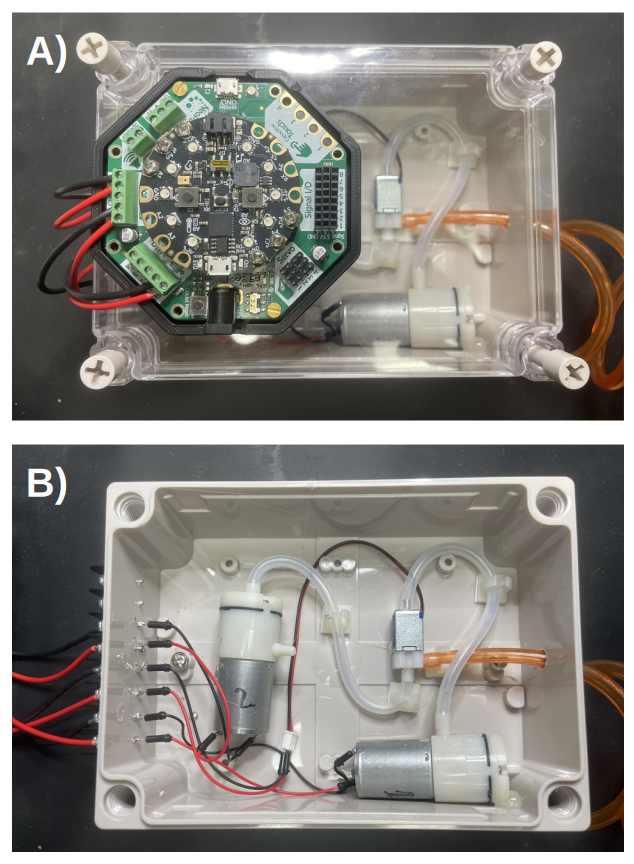}
    \caption{Housing for phantom electronics and mechanical hardware. (A) The housing shown with the CRICKIT in place. (B) The motors, tubes, and wire harness shown in the enclosure. Each feed through has strain relief to provide robustness to the design. Additionally, the housing is modular and may be modified or added onto using standoffs, as shown, for other custom uses.}
    \label{fig:housing}
\end{figure}

We built a housing for the electrical and mechanical components of the phantom, shown in Figure \ref{fig:housing}. This consists of a 3D printed external mount for the CPX/CRICKIT (design file \texttt{mount-\allowbreak crickit-\allowbreak noinserts.stl}), and a container for the pumps and pneumatic valve. The plastic container protects the controller hardware from mechanical impact and liquid intrusion. The container is plastic and easily machinable, allowing users to customize their electrical and pneumatic interface to the container with components which are most readily available.

In summary, the device described can support MRI development, particularly at ULF, by:

\begin{itemize}
\item Allowing for testing of image acquisition, pulse sequences, and reconstruction algorithms in a phantom with periodic motion mimicking that of human heart ventricles.
\item Providing a low-cost, open-source system for easy customization of simulated cardiac cycles.
\item Producing dynamic motion without the introduction of EMI, a particularly valuable feature for unshielded MRI applications.
\item Providing a robust, compact control system ideal for teaching labs where students may gain hands-on exposure to medical imaging concepts.
\end{itemize}

\section{Design files summary}

A complete set of the design files required for construction and operation of the phantom is deposited in the Zenodo repository. \texttt{simple\_\allowbreak cardiac\_\allowbreak cycle.ino} is the Arduino sketch used to control the phantom's motion. \texttt{balloon\_\allowbreak phantom\_\allowbreak V1.2.stl} is the CAD design file for the phantom housing, and \texttt{balloon\_\allowbreak phantom\_\allowbreak V1.2.f3d} is the associated Autodesk Fusion Archive (Autodesk, Inc., San Francisco, CA, USA) file with edit history and metadata, provided as the editable source format. \texttt{mount-\allowbreak crickit-\allowbreak noinserts.stl} is a CAD design file for the component used to mount the CPX/CRICKIT onto the motor enclosure --- this was repurposed with no changes from the demo \url{https://learn.adafruit.com/mount-for-crickit}, where it is available under the CC BY 4.0 license. \texttt{retrospective\_\allowbreak gating\_\allowbreak recon.m} is the MATLAB code used to perform the retrospectively gated image reconstruction described in Section \ref{sec:validation}.

\vskip 0.1cm
\tabulinesep=1ex
\begin{tabu} to \linewidth {|X[1.5,l]|X[l]|X[1.2,l]|X[1.5,l]|}
\hline
\textbf{Design filename} & \textbf{File type} & \textbf{Open source license} & \textbf{Location of the file} \\\hline
\texttt{simple\_\allowbreak cardiac\_\allowbreak cycle.ino} & Arduino source file & CC BY 4.0 & \href{https://doi.org/10.5281/zenodo.17653758}{Zenodo repository} \\\hline
\texttt{balloon\_\allowbreak phantom\_\allowbreak V1.2.f3d} & Autodesk Fusion CAD source file (editable) & CC BY 4.0 & \href{https://doi.org/10.5281/zenodo.17653758}{Zenodo repository} \\\hline
\texttt{balloon\_\allowbreak phantom\_\allowbreak V1.2.stl} & CAD file for 3D printing & CC BY 4.0 & \href{https://doi.org/10.5281/zenodo.17653758}{Zenodo repository} \\\hline
\texttt{mount-\allowbreak crickit-\allowbreak noinserts.stl} & CAD file for 3D printing & CC BY 4.0 & \href{https://doi.org/10.5281/zenodo.17653758}{Zenodo repository} \\\hline
\texttt{retrospective\_\allowbreak gating\_\allowbreak recon.m} & MATLAB source file & CC BY 4.0 & \href{https://doi.org/10.5281/zenodo.17653758}{Zenodo repository} \\\hline
Supplemental Video S1 & Annotated video of phantom motion & CC BY 4.0 & Available with the article \\\hline
\end{tabu}

Video analysis code used to extract the fluid level and balloon volume timecourses from Supplemental Video S1 is additionally maintained at \url{https://github.com/jonbmartin/cmr-phantom-analysis}.

\section{Bill of materials summary}

Table \ref{tab:bom} provides a complete list of materials for the construction of the phantom along with actual costs. All materials and components necessary to build and operate the phantom are provided, including the suppliers. Designators in the first column correspond to the labels used in the pneumatic schematic (Fig.\ \ref{fig:schematic}) and the build instructions below. The most expensive material is the mineral oil used to fill the phantom, a commonly used material but not one which is strictly necessary for this application. More affordable filling liquids could be utilized, such as a saline or copper sulfate solution. Excluding the filling liquid, the phantom cost was \$187.13. Using mineral oil increases the total cost to \$219.25. All costs are in 2026 USD.

\vskip 0.2cm
\begin{longtable}{L{2.5cm} L{4.9cm} C{0.7cm} C{1.7cm} C{1.7cm} L{2.1cm} L{1.8cm}}
\caption{Complete bill of materials for phantom construction. All costs are in 2026 USD.}
\label{tab:bom}\\
\hline
\textbf{Designator} & \textbf{Component} & \textbf{No.} & \textbf{Unit cost (USD)} & \textbf{Total cost (USD)} & \textbf{Source of materials} & \textbf{Material type} \\
\hline
\endfirsthead
\hline
\textbf{Designator} & \textbf{Component} & \textbf{No.} & \textbf{Unit cost (USD)} & \textbf{Total cost (USD)} & \textbf{Source of materials} & \textbf{Material type} \\
\hline
\endhead
\hline
\endfoot
CRICKIT & Adafruit CRICKIT & 1 & \$29.99 & \$29.99 & Adafruit & Non-specific \\\hline
CPX & Adafruit Circuit Playground Express & 1 & \$39.95 & \$39.95 & Adafruit & Non-specific \\\hline
T-connector & T-connector for 3 mm ID silicone tubing & 1 & \$4.50 & \$4.50 & Adafruit & Polymer \\\hline
Silicone tubing & 3 mm ID silicone tubing & 1 & \$7.95 & \$7.95 & Adafruit & Polymer \\\hline
Air Valve & 6V air valve with 2-pin JST XH connector & 1 & \$2.95 & \$2.95 & Adafruit & Non-specific \\\hline
Air Pump & Air pump and vacuum DC motor, 2.5 LPM & 2 & \$7.95 & \$15.90 & Adafruit & Non-specific \\\hline
Power supply & 5V 2A switching power supply & 1 & \$7.95 & \$7.95 & Adafruit & Non-specific \\\hline
Epoxy adhesive & PlasticWeld epoxy & 1 & \$7.52 & \$7.52 & J-B Weld & Polymer \\\hline
Electrical tape & 33+SUPER-3/4X44FT vinyl electrical tape & 1 & \$10.00 & \$10.00 & DigiKey & Polymer \\\hline
PLA filament & Polymaker Polyterra PLA filament & 1 & \$20.99 & \$20.99 & DigiKey & Polymer \\\hline
Plasti Dip & Plasti Dip multi-purpose spray-on rubber coating, 11 oz. & 1 & \$12.17 & \$12.17 & U.S. Plastic Corp. & Polymer \\\hline
Plastic Enclosure & Large plastic project enclosure, weatherproof with clear top & 1 & \$19.95 & \$19.95 & Adafruit & Polymer \\\hline
Barrier Block & 8-circuit 8.26 mm barrier block connector & 1 & \$2.82 & \$2.82 & DigiKey & Non-specific \\\hline
Balloon & 3 inch inflatable latex balloon & 1 & \$4.49 & \$4.49 & Walmart & Polymer \\\hline
Mineral oil & 1 pt.\ white mineral oil & 2 & \$16.06 & \$32.12 & McMaster-Carr & Organic \\\hline
\multicolumn{4}{l}{\textbf{Total}} & \textbf{\$219.25} & & \\\hline
\multicolumn{4}{l}{\textbf{Total excluding mineral oil}} & \textbf{\$187.13} & & \\\hline
\end{longtable}

\section{Build instructions}

\subsection{3D printing and assembly}
Design files for the 3D printed phantom structure are found in the Zenodo repository (\url{https://doi.org/10.5281/zenodo.17653758}). The phantom (\texttt{balloon\_\allowbreak phantom\_\allowbreak V1.2.stl}) was printed on a filament extrusion printer using PLA filament. Depending on the available printing volume, the phantom may be split into cylindrical segments and epoxied together after printing. To create a liquid-tight structure and prevent mineral oil from leaking, the 3D printed structure was sprayed with two coats of Plasti-Dip rubberizing spray (McMaster-Carr PN 9560T4, Color: Semigloss White).

The silicone tubing was inserted approximately 2 inches into the balloon to prevent blockage of the tubing by the narrow neck of the balloon. The balloon opening with inserted tubing was sealed with epoxy to produce a water-tight cavity. The tubing was then fastened with epoxy to the bottom of the 3D printed phantom structure, allowing the natural buoyancy of the balloon to suspend it in the column of liquid. The tubing was run up and out of the housing through the space between the inner and outer 3D printed tubes (see Fig.\ \ref{fig:function}), to prevent the tubing from interfering with the motion of the balloon. The phantom housing was finally filled with mineral oil to within approximately 1.5 inches of the top of the outer tube, to allow space for the liquid level to rise and fall.

\subsection{Pneumatic circuit assembly and housing construction}
An overview of the pneumatic system used to control the balloon's inflation is shown in Figure \ref{fig:schematic}, and a picture of the constructed pneumatic system is shown in Figure \ref{fig:housing}B. First, 8 evenly spaced 5/64" holes (left side, Fig.\ \ref{fig:housing}B) and 1 1/4" hole (right side, Fig.\ \ref{fig:housing}B) were drilled into the plastic enclosure for the electrical connections and the pneumatic line respectively. An 8-circuit barrier block was inserted into the 5/64" holes and fixed in place with epoxy. Both air pumps and the air valve were fixed to the bottom of the enclosure with epoxy. Silicone tubing was used to connect both air pumps to the air valve. Because the air pumps used are unidirectional, the intake of one pump and outtake of the other pump were connected to the valve's side inputs. Silicone tubing connecting to the balloon was attached to the middle common connection. The electrical connections for both pumps and the valve were soldered to the interior side of the barrier block. Pump ``1'' (bottom right, Fig.\ \ref{fig:housing}B) and pump ``2'' (left, Fig.\ \ref{fig:housing}B) were connected to the CRICKIT's ``1'' and ``2'' DC motor outputs respectively. The air valve was connected to the CRICKIT's high-current drive output. The CRICKIT/CPX was inserted into the custom 3D printed mount (\texttt{mount-\allowbreak crickit-\allowbreak noinserts.stl}), which was attached with epoxy onto the top of the plastic enclosure.

\subsection{Safety considerations during construction}
Construction requires drilling, soldering, and the use of two-part epoxy and an aerosol rubberizing spray. Eye protection should be worn while drilling the enclosure. Both the epoxy and the Plasti-Dip spray should be applied in a well-ventilated area, and the manufacturer's safety data sheets should be consulted. The 3D printed housing must be fully cured and leak-tested with water before being filled with mineral oil, and before being brought into the scanner room, to avoid spilling oil onto RF coil hardware. Note that the control electronics, pumps, and valve contain ferromagnetic components and are not MR-safe; see Section \ref{sec:operation} for the associated siting requirements.

\section{Operation instructions}
\label{sec:operation}

\begin{figure}[htbp]
    \centering
    \includegraphics[width=0.5\linewidth]{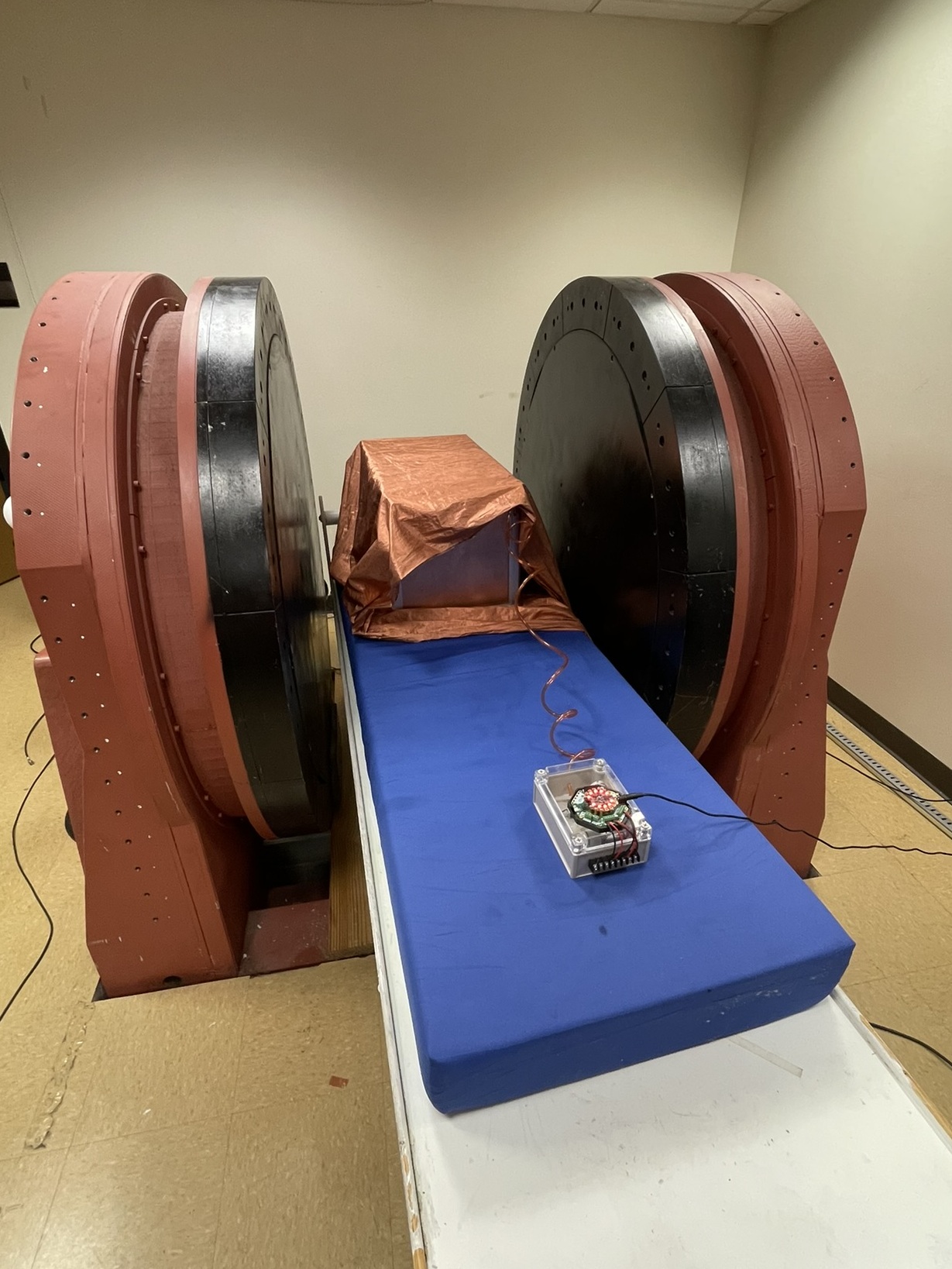}
    \caption{Control system housing near the 47.5 mT biplanar permanent magnet MRI. For reference, the coil and phantom are inside the local copper shielded box placed at isocenter.}
    \label{fig:with_magnet}
\end{figure}

The 3D printed phantom housing, containing the balloon and mineral oil, is placed into an RF coil within the magnet. If the system on which imaging will be performed includes a shielded room, the pneumatic line may be run out of the scanner room via an interface panel. For our unshielded low-field magnet, this was not necessary and the enclosure was able to be positioned close to the magnet (see Fig.\ \ref{fig:with_magnet}). In general, as the hardware used to drive the pneumatic phantom has some ferromagnetic components and is not MR-safe, it is recommended that the control housing be kept outside of the MRI system's 5 Gauss line, both to ensure safety and for the successful operation of the phantom. In the 47.5 mT system used in this study, the magnetic field drops off quickly enough to allow the enclosure to be placed at the end of the patient bed. If a TTL trigger is desired for control of sequences, this signal is produced by the CRICKIT signal I/O channel \#1 (see Fig.\ \ref{fig:schematic}) and may be connected to triggering hardware.

Control of the pneumatic system is performed using an Arduino script; an example script (\texttt{simple\_\allowbreak cardiac\_\allowbreak cycle.ino}) is included in the Zenodo repository. The duty cycle and throttle of the pumps may be modified to control the motion dynamics. To maximize repeatability and ensure that the balloon does not over-inflate, the deflating cycle is made longer than the inflating cycle, which results in the balloon fully deflating in each cycle. This was achieved with a 40\% duty cycle. The NeoPixel LEDs on the CPX alternate between green and red to indicate whether the balloon is inflating or deflating. This allows for visual monitoring of the balloon's inflation cycle, which is not visible from within the 3D printed housing. The action of the phantom may be terminated by pressing the CPX D4 input button. When pressed, the phantom immediately switches from its inflate/deflate cycle to deflating the balloon, and after complete deflation the program terminates.

\subsection{Operational safety hazards}
Two hazards should be noted. First, the control electronics, DC pumps, and solenoid valve contain ferromagnetic components and must not be brought inside the 5 Gauss line; they constitute a projectile risk at higher field strengths. Second, the balloon is inflated without pressure or volume feedback (see Section \ref{sec:limitations}), so the duty cycle must be configured such that the deflation phase is at least as long as the inflation phase to prevent over-inflation and possible rupture of the balloon inside the housing. We recommend bench-testing any modified timing parameters outside the magnet before use. Mineral oil is a slip hazard if spilled; the phantom lid should be fitted during transport.

\section{Validation and characterization}
\label{sec:validation}

\subsection{Video validation of phantom motion}
To provide independent characterization of the motion dynamics of the phantom, a 60 second video recording was made of the top liquid surface within the phantom while the phantom operated. The CRICKIT/CPX was included in frame, so that the flashing LED lights could be used to calibrate the pneumatic delay between TTL trigger and beginning of the inflationary cycle. A ruler was inserted into the phantom in the field of view to calibrate the camera pixel resolution to a known length. Fluid level was manually annotated on the video in each frame. Balloon volume at each timestep was derived from the known size of the phantom holder using Archimedes' principle. Frames in which the red CRICKIT/CPX LED turned on were manually annotated as timepoints at which the TTL signal activated. The annotated version of the recorded video, with extracted liquid height, is included as Supplemental Video S1. Video analysis code may be found in the following GitHub repository: \url{https://github.com/jonbmartin/cmr-phantom-analysis}.

\begin{figure}[htbp]
    \centering
    \includegraphics[width=1\linewidth]{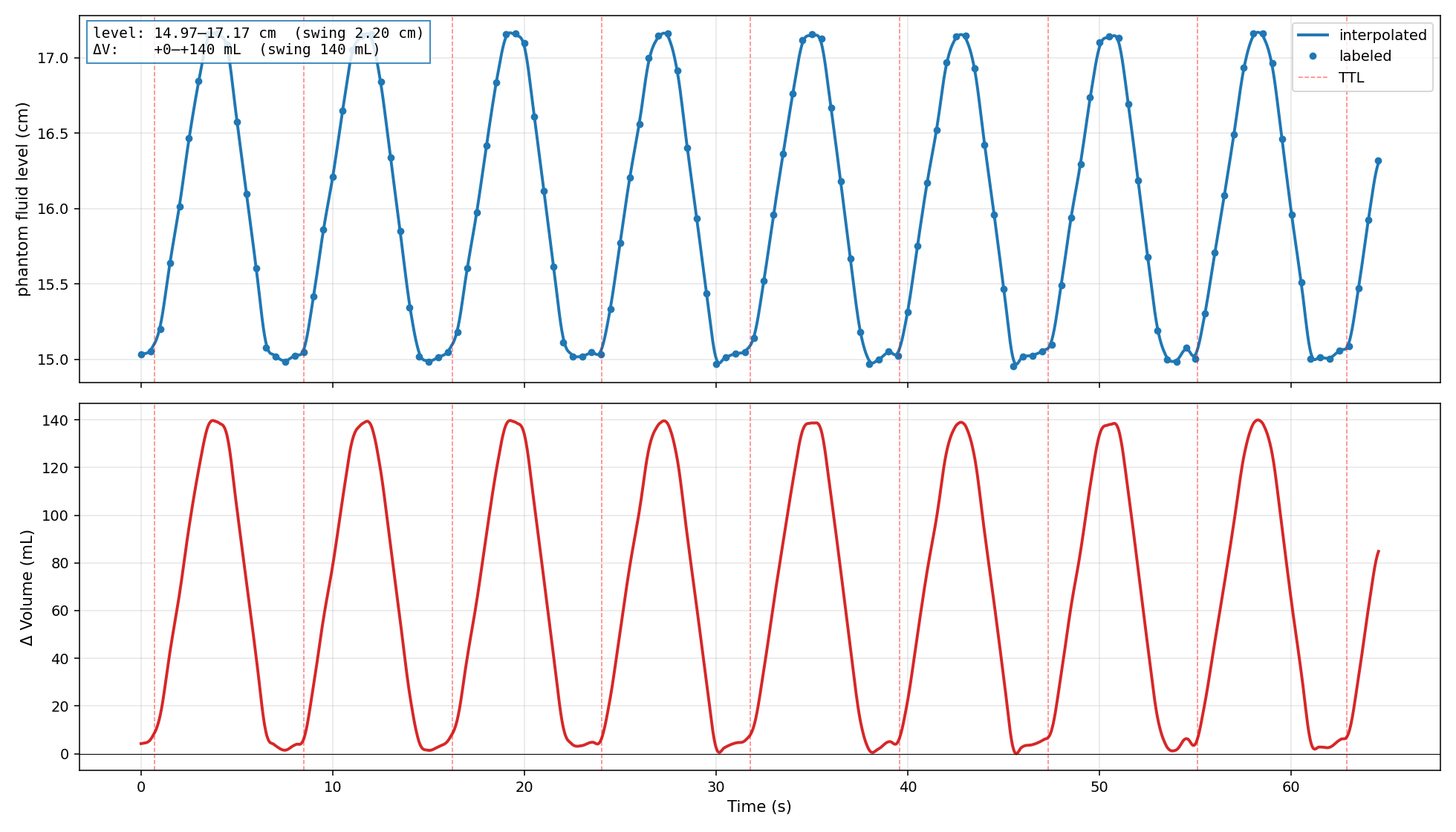}
    \caption{Motion dynamics of the phantom, derived from video recording. (Top) Timecourse of fluid level in phantom holder, with TTL trigger marked. (Bottom) Inferred change in phantom volume from baseline.}
    \label{fig:level_volume}
\end{figure}

Figure \ref{fig:level_volume} shows the timecourse of video-derived phantom fluid level and the corresponding derived change in balloon volume. The motion cycle was observed to be extremely stable. Over the 8 measured cycles, peak-to-peak interval was 7.742 $\pm$ 0.043 s (0.6\% CoV). Maximum balloon $\Delta$V per cycle was 140.0 $\pm$ 3.2 mL (2.3\% CoV). From the video, it was determined that pneumatic temporal delay was small, with a delay from LED/TTL activation to inflation of 10 $\pm$ 16 ms. This is consistent with the programmed sequence, which included an intentional 10 ms delay between TTL and pump activation. Thus additional pneumatic temporal delay was negligible.

\subsection{Ultra-low field MR imaging}
To demonstrate a practical use case, we acquired images on an ULF MRI scanner based on a 47.5 mT permanent magnet (Sigwa Company, LLC, Acton, MA, USA). A 2D balanced steady-state free precession (bSSFP) retrospectively gated cine imaging sequence was acquired. Testing was performed in a slow cardiac cycle, while the phantom was operating at a 40\% duty cycle with a period of 7 s. The 2D bSSFP sequence parameters were TR/TE = 13/6.5 ms, bandwidth = 32 kHz, RF flip angle = 70$^\circ$, matrix size = 128 $\times$ 64, dx = 3.0 mm, dy = 3.5 mm, time resolution = 0.636 s over 11 phases, total scan time = 28 min. To provide data on repeatability, the 28 minute scan was divided into 4 seven-minute data subsections, each of which were used to derive an independent motion cycle. To reconstruct the image time series via retrospective gating \cite{lanzer_cardiac_1984, lenz_retrospective_1989}, the CRICKIT-generated TTL signal was recorded using a PicoScope 4000 Series digital oscilloscope (Pico Technology Ltd., St. Neots, United Kingdom), and the acquired data were segmented into 11 phases using a retrospective data sorting approach wherein phase encodes were spaced over an interval greater than the R-R interval and retrospectively assigned to their nearest bin (e.g. method `C', Ref.\ \cite{lenz_retrospective_1989}). MATLAB code performing the reconstruction is available in the Zenodo repository.

\begin{figure}[htbp]
    \centering
    \includegraphics[width=0.66\linewidth]{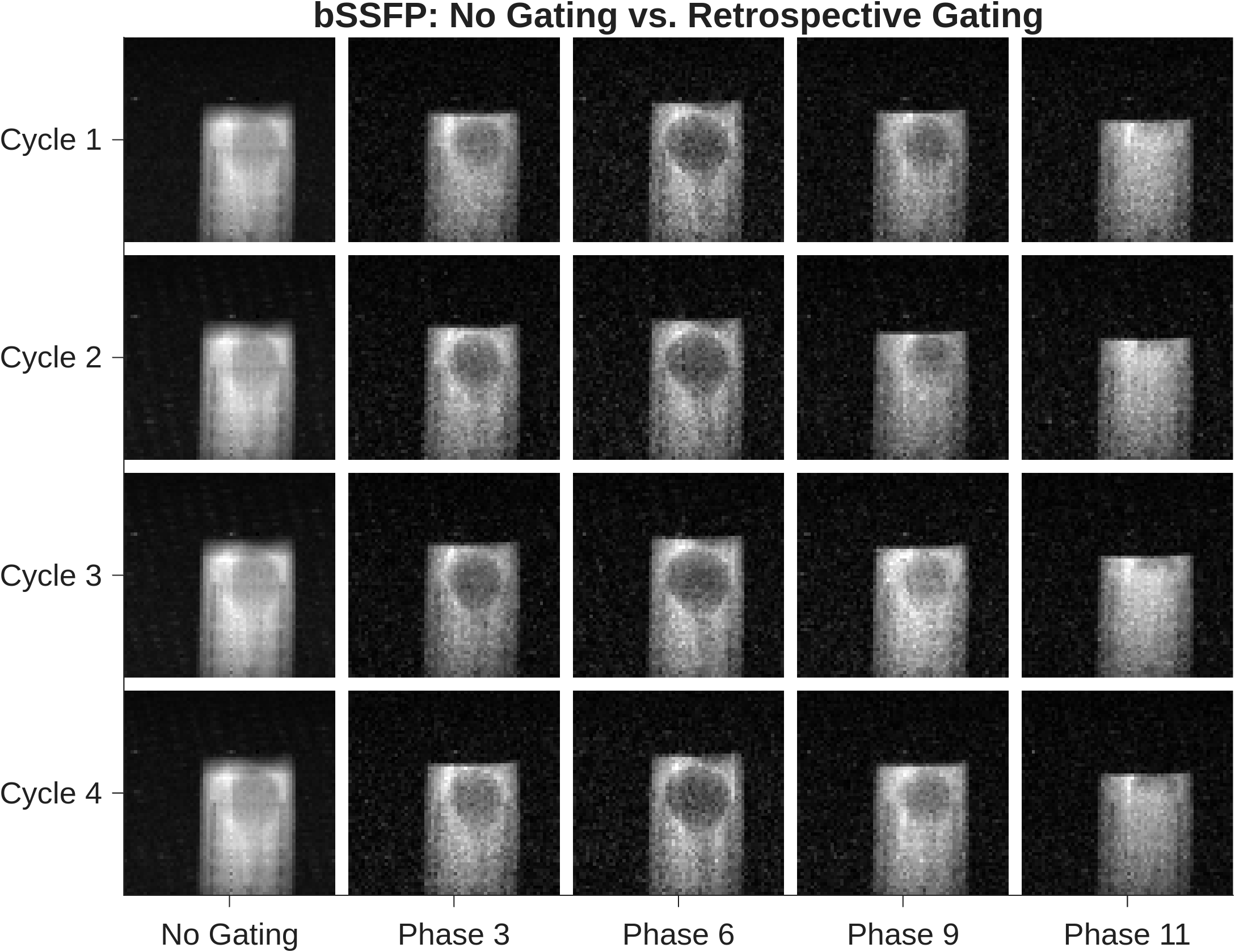}
    \caption{Images from 0.047 T bSSFP acquisition across 4 cycles, with and without retrospective gating. The leftmost column shows the ungated acquisition. The remaining columns show images reconstructed across representative phases with retrospective gating.}
    \label{fig:47mt}
\end{figure}

\begin{figure}[htbp]
    \centering
    \includegraphics[width=1\linewidth]{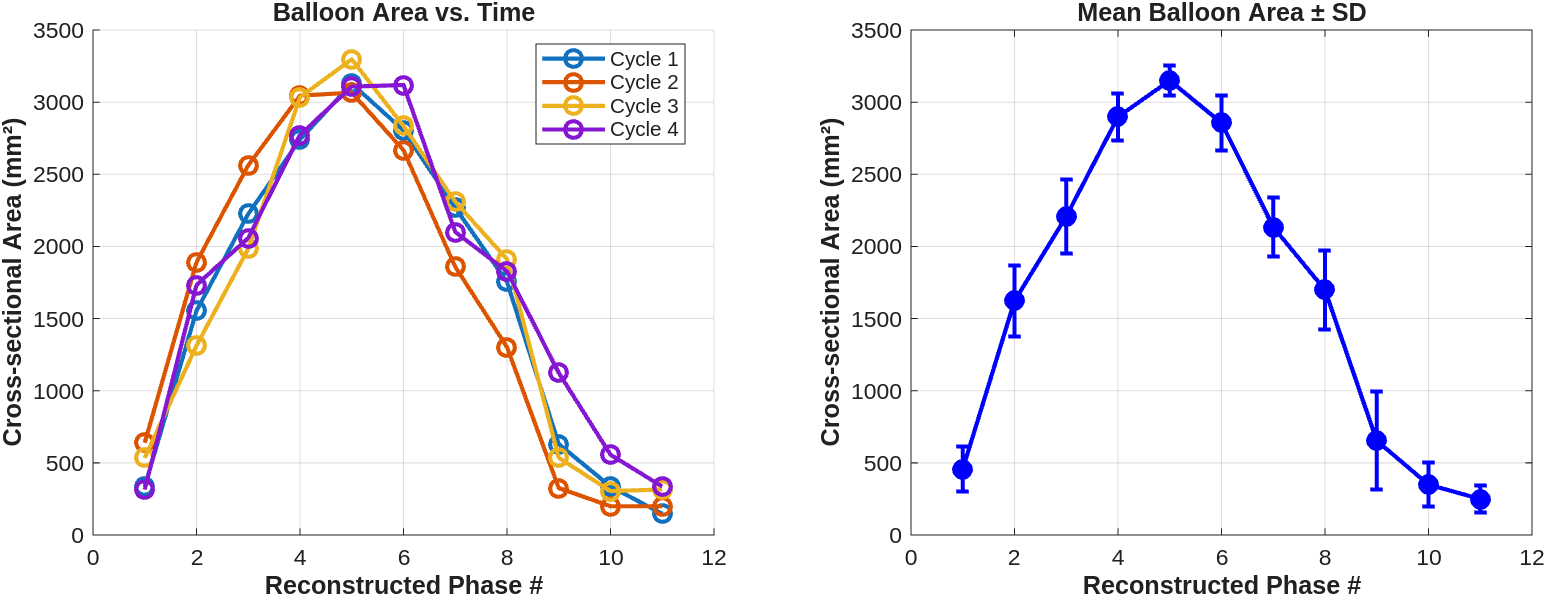}
    \caption{(L) Cross-sectional balloon area across reconstructed phases, from 0.047 T bSSFP images (Fig.\ \ref{fig:47mt}). (R) Mean and standard deviation of area measured across the same reconstructed phases.}
    \label{fig:repeatable}
\end{figure}

Reconstructed ULF MR images of the phantom in operation are shown in Figure \ref{fig:47mt}. If all acquired data are reconstructed together without retrospective segmentation, the motion of the phantom over the slow cardiac cycle results in blurring of the image and loss of contrast between the balloon and the surrounding mineral oil. Segmenting the data to the balloon's inflation cycle allows for a time-resolved reconstruction in which periodic changes in the mineral oil level and volume of the balloon over time are discernible. A simple estimate of the cross-sectional area of the balloon across phases was produced by manually segmenting the attenuated region within the phantom using MATLAB's ROI tool. A plot of this cyclic variation in cross sectional area is shown in Fig.\ \ref{fig:repeatable} for all 4 acquired repetitions. MRI-derived changes in balloon cross-sectional area over time was found to have excellent repeatability, with an intraclass correlation coefficient of 0.958. Mean maximum cross sectional area was 3150 mm$^2$, with a CoV of 3.22\%. There was close agreement between the MR-derived cross sectional area measurement and that inferred from the video measurement. Assuming a spherical balloon at maximum inflation, a maximum volume of 140 mL implies a cross-sectional area of 3260 mm$^2$, a difference of only 3.37\%. These imaging results demonstrate the feasibility of using this phantom as a platform for evaluating image acquisition and reconstruction innovations, and retrospective image analyses.

\subsection{Electromagnetic interference}
EMI can severely impact image quality in MRI. EMI mitigation is particularly important in unshielded MRI systems, which are in use mainly at low field strengths \cite{vliem_design_2026, lena_subject_2026, srinivas_external_2022}. These are motivated primarily by potential for reductions in cost and increased system portability for point-of-care applications. Although mitigation of EMI is possible, it would be desirable for the developed phantom to be able to operate in close proximity to an unshielded system without generating EMI, the removal of which complicates imaging. Although the phantom is based on pneumatics and DC motors, which are known to produce minimal EMI, the large number of active electrical and mechanical components, coupled with the potential use in unshielded ULF MRI systems, warranted an evaluation of the EMI produced during operation.

A free induction decay (FID) measurement was performed with the ULF MR system's spectrometer over a bandwidth of 32 kHz centered at the Larmor frequency. This measurement was performed repeatedly in 5 second intervals over 280 s, with the phantom present in the volume coil and the electronics housing positioned as in Fig.\ \ref{fig:with_magnet} but powered off for 140 s. At this point, the phantom was switched on for another 140 s. To provide comparison to a high-EMI environment, this experiment was repeated with the phantom present but inactive for another period of 280 s but with a known EMI source activated at 140 s. The known EMI source was produced by a multi-turn loop coil tuned to the 47.5 mT Larmor frequency external to the MR system and driven with a RIGOL DG1022 function generator (RIGOL Technologies, Inc., Portland, OR, USA) producing a 5 Vpp frequency sweep centered about the Larmor frequency.

\begin{figure}[htbp]
    \centering
    \includegraphics[width=1\linewidth]{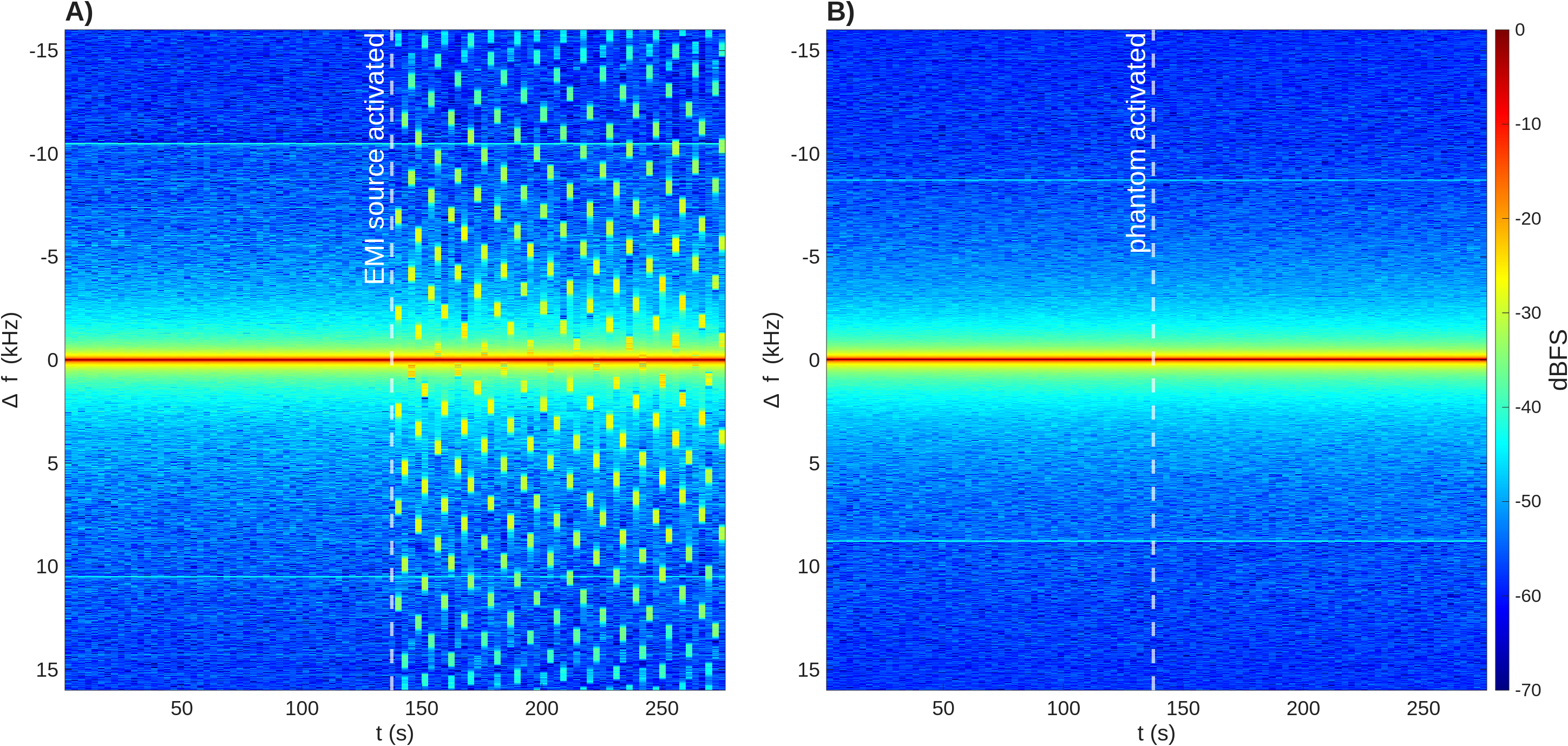}
    \caption{Spectrogram showing MR FID and EMI measured by the 47.5 mT imaging system while A) a known EMI source is turned on in the scan room and B) while the phantom is turned on, with the actuators and electronics situated only a few feet away from the system. Full scale signal value at $\Delta f$ = 0 Hz is the mineral oil MR signal produced by FID.}
    \label{fig:EMI}
\end{figure}

The recorded spectrograms of both experiments are shown in Figure \ref{fig:EMI}. Unlike in the case of the known source of EMI, we did not detect EMI in the case of the motion phantom activation, verifying that the pneumatic phantom and its associated electronics are suitable for unshielded low-field MRI systems. To quantify this, we computed band-integrated spectral power within a $\pm$ 2 kHz signal band and an off-resonance noise reference band ($|\Delta f| > 8$ kHz) for each measurement. Upon activation of the balloon phantom, the integrated signal-band power changed by $-0.07 \pm 0.01$ dB. The noise floor was unchanged at $-55.7$ / $-55.8$ dBFS (before/after), and SNR remained constant at 36.5 dB, confirming no measurable EMI contribution from the phantom. In contrast, the positive-control fast-sweep EMI source raised the off-resonance noise floor by 9.8 dB ($-54.5$ to $-44.7$ dBFS), reducing SNR from 35.4 to 25.8 dB. We also note that although the mechanical and electrical components of the phantom were in close proximity to the low field magnet ($<$ 3 ft.\ distance, shown in Fig.\ \ref{fig:with_magnet}), the pumps and motors were able to continue to function in the fringe of the magnetic field.

\subsection{Summary of measured capabilities and limitations}

Measured performance of the hardware, as characterized above:

\begin{itemize}
\item \textbf{Cycle period:} user-programmable; 7.742 $\pm$ 0.043 s demonstrated (0.6\% coefficient of variation over 8 consecutive cycles).
\item \textbf{Stroke volume:} 140.0 $\pm$ 3.2 mL change in balloon volume per cycle (2.3\% CoV), measured by video analysis.
\item \textbf{Cross-sectional area at maximum inflation:} 3150 mm$^2$ by MRI (3.22\% CoV), in 3.37\% agreement with the volume-derived estimate of 3260 mm$^2$.
\item \textbf{Repeatability under MRI:} intraclass correlation coefficient of 0.958 for the area-versus-phase curve across 4 independent 7-minute acquisitions within a 28-minute session.
\item \textbf{Trigger latency:} 10 $\pm$ 16 ms from TTL assertion to onset of inflation, consistent with the 10 ms delay programmed in firmware; additional pneumatic latency is therefore negligible.
\item \textbf{Duty cycle:} 40\% used here; programmable, subject to the constraint that deflation must be at least as long as inflation to avoid over-inflation.
\item \textbf{Maximum pump flow rate:} 2.5 LPM per pump.
\item \textbf{Electromagnetic compatibility:} no measurable EMI contribution during operation (signal-band power change $-0.07 \pm 0.01$ dB, noise floor unchanged within 0.1 dB, SNR constant at 36.5 dB) with the control housing $<$ 3 ft.\ from an unshielded 47.5 mT magnet.
\item \textbf{Filling liquid relaxation times:} white mineral oil $T_1$ = 114.153 $\pm$ 1.978 ms and $T_2$ = 105.087 $\pm$ 2.586 ms at 47.5 mT (see Supplementary Information).
\end{itemize}

Corresponding limitations:

\begin{itemize}
\item The balloon is air-filled, so there are no spins within it; bright-blood contrast and physiological flow cannot be simulated.
\item There is no closed-loop pressure or volume feedback, so over-inflation is prevented by timing rather than by measurement.
\item The phantom must be oriented vertically because the housing is lidded but not sealed.
\item Mineral oil $T_1$/$T_2$ are not matched to myocardium or lung parenchyma.
\item Performance has been validated only at 47.5 mT, at the relatively coarse spatiotemporal resolution available on that system.
\item Long-term durability and fatigue life of the latex balloon have not been characterized.
\end{itemize}

These points are discussed further in Section \ref{sec:limitations}.

\section{Limitations}
\label{sec:limitations}

The motion phantom has several limitations and considerations that should be made prior to its use. Many of these are consequences of the steps taken to keep the phantom low cost and simplify its design, assembly, and use. Contrast is currently generated by filling a balloon surrounded by mineral oil with air. A lack of spins inside the balloon prevents the production of bright-blood contrast or simulation of physiological flow into or out of the balloon \cite{evans_cardiac_1988}. Thus, this phantom is most appropriate for prototyping imaging sequences (and the associated reconstruction/postprocessing) that do not assess blood flow, such as 4D flow imaging. Relatively minor modifications could extend the design to incorporate a fluid-filled balloon, for example by replacing the pneumatic air pumps with peristaltic pumps. An advantage of using air pumps is that the flow rates are typically higher than those provided by peristaltic pumps (our air pumps operate at 2.5 liters per minute maximum), allowing the balloon to be inflated to larger volumes more quickly than could be achieved by a typical inexpensive peristaltic pump.

Another limitation of the phantom as described are the $T_1$ and $T_2$ values of the mineral oil filling liquid. Mineral oil was chosen as it is a common phantom filling liquid due to its short $T_1$, stability, relative affordability, and high availability. However, mineral oil $T_1$ and $T_2$ values are not similar to those of the most relevant moving tissues, such as myocardium or lung parenchyma. We measured the white mineral oil $T_1$ and $T_2$ at 0.047 T to be 114 ms and 105 ms respectively (see Supplementary Information). $T_1$ and $T_2$ mapping of native myocardium has not been performed at this field strength; at 0.35 T, the lowest field strength commonly employed for cardiac MRI, myocardium is reported to have a longer $T_1$ of 564 ms and shorter $T_2$ of 59 ms \cite{campbell-washburn_cardiac_2024}. Nickel chloride or copper sulfate solutions could be alternative filling liquids that more closely emulate tissues of interest. To avoid degradation of the balloon latex due to exposure to transition metal solutions, in these cases a compliant silicone rubber balloon would be used instead (e.g.\ YJRB Custom Silicone Rubber Balloons, Yejia Optical Technology). Such balloons are commonly used for biomedical applications requiring repeatable inflation with excellent material elasticity, for example for catheterization. Such a balloon is recommended as a more durable alternative to the current latex balloon.

The phantom does not have a mechanism for direct feedback on volume or pressure inside the balloon, which can potentially lead to over-inflation. This also makes validation of measured volumes more challenging, although the incorporation of an inner 3D printed tube to constrain the expansion does provide a known reference cross-sectional area at maximum inflation. Future iterations of the phantom could include a liquid level sensor, such as a mechanical float, to provide an estimation of the displaced volume through the simulated cardiac cycle. To avoid the potential for mechanical failure due to over-inflation and to improve cycle repeatability, our control logic currently implements a deflationary cycle that is longer than the inflationary cycle, ensuring that the balloon is entirely deflated and begins each inflationary cycle with the same initial volume. We have not characterized the long-term durability and fatigue life of the phantom. Latex balloons in general have extremely high elasticity if they are not over-inflated. We observed a high degree of compliance and repeatable inflation from the existing latex balloon construction over a relatively long 30 minute imaging session (see Fig.\ \ref{fig:repeatable}).

We stress that thus far the performance of this phantom has only been validated on an ultra-low field 47.5 mT MRI system with relatively limited spatiotemporal resolution. Cycle-to-cycle variations in phantom motion were not detected to be deleterious to low resolution imaging in this work. However it is possible that to translate the phantom for higher-resolution imaging applications some modifications, such as additional contacts tethering the balloon, may be necessary. The need for extremely low cost phantoms is generally lessened in the context of high cost, higher field strength systems, where the cost of an individual phantom becomes nearly negligible relative to the cost of building, testing, and maintaining a MRI system. In these settings, a higher-cost commercial phantom could be more appropriate.

The phantom currently must be positioned vertically in the magnet, as shown in Fig.\ \ref{fig:function}, due to the lidded but unsealed top of the 3D printed housing. This could provide a challenging geometric constraint for some RF coils, although the height of the phantom approximates average human anterior-posterior chest height \cite{beesems_force_2015} and can fit in a typical head coil. A future iteration of this device featuring a sealed phantom which may be positioned in arbitrary orientations is being developed.

Despite these limitations, the phantom overall produces a regular and repeatable motion, which agreed with independent validation by video analysis. Variation in maximum cross-sectional area in bSSFP measurement varies by only 3.22\%. This is below the level of variability in morphological measurement seen, for example, in repeat measurements in a single image session in conventional high-field cardiac MRI \cite{danilouchkine_operator_2005}.

\section{Conclusion}

Imaging phantoms that substitute for human participants during the development or evaluation of imaging techniques are valuable for the development of cardiac and lung imaging. Existing phantoms are expensive, bulky, have limited programmability, or are inappropriate for the unshielded MR environment. We developed a low-cost, simple, pneumatic phantom to overcome these challenges. The phantom is particularly suited to the needs of ULF MRI, and represents to our knowledge the first motion phantom imaged at ULF.

The phantom introduced here is the first open-source motion phantom we are aware of that is controlled with an embedded microcontroller system. The adaptability and customization potential of microcontrollers makes this phantom well suited for teaching and demonstrations. It also enables rapid modification, customization, and expansion of the system by more advanced users for more complex imaging scenarios.

The low cost of this phantom makes it particularly attractive for ULF MRI, as a notable motivation for adopting ULF is that dramatically more affordable systems become possible \cite{sarracanie_low-cost_2015}. As interest in body imaging in general and cardiac imaging in particular increase at low \cite{campbell-washburn_cardiac_2024} and ultra-low fields \cite{zhao_whole-body_2024, lau_ultra-low-field_2025} we hope that this phantom may serve as a useful tool for methods development.

\vskip 0.3cm
\noindent
\textbf{Ethics statements}\\

\noindent
This work did not involve human subjects, human data, or animal experiments. All measurements reported were made on the inanimate phantom described in this article.\\

\noindent
\textbf{CRediT author statement}\\

\noindent
\textbf{Jonathan B. Martin}: Writing -- original draft preparation, Conceptualization, Methodology, Investigation, Software, Visualization, Funding acquisition. \textbf{Millie A. Barron}: Writing -- review \& editing, Investigation. \textbf{Charlotte R. Sappo}: Writing -- review \& editing, Conceptualization, Methodology, Visualization, Resources, Funding acquisition.\\

\noindent
\textbf{Declaration of competing interest}\\

\noindent
The authors declare that they have no known competing financial interests or personal relationships that could have appeared to influence the work reported in this paper.\\

\noindent
\textbf{Declaration of generative AI and AI-assisted technologies}\\

\noindent
Claude Code using the Sonnet 4.6 model (Anthropic, PBC, San Francisco, CA, USA) assisted in the writing of the video analysis code used in Section \ref{sec:validation}. The authors reviewed and verified all generated code and take full responsibility for the content of this article.\\

\noindent
\textbf{Acknowledgements}\\

\noindent
Funding: This work was supported by the National Institutes of Health, National Heart Lung and Blood Institute [grant number K25HL183904], the Vanderbilt Institute for Clinical and Translational Research VR73376, and the American Association of University Women (AAUW) Award Number G-2025-14902. Its contents are solely the responsibility of the authors and do not necessarily represent official views of the National Institutes of Health, Vanderbilt, or AAUW.\\

\noindent
\textbf{Data and design file availability}\\

\noindent
All design files, source code, and CAD models for the phantom are openly available in the Zenodo repository at \url{https://doi.org/10.5281/zenodo.17653758} under the CC BY 4.0 license. Video analysis code is available at \url{https://github.com/jonbmartin/cmr-phantom-analysis}.

\bibliographystyle{unsrt}
\bibliography{references}

\newpage
\section*{Supplementary Information}

\subsection*{S1. Relaxometry of the mineral oil filling liquid}

Relaxometry measurements were performed to characterize the $T_1$ and $T_2$ of the white mineral oil used in the phantom on our 47.5 mT system. $T_1$ measurements were made using a non-imaging inversion recovery sequence. The inversion pulse was a hyperbolic secant pulse designed using MRiLab v1.3 in MATLAB \cite{liu2016fast} (\url{https://leoliuf.github.io/MRiLab/}) and was followed by a 90$^\circ$ hard pulse before readout. A spoiler gradient after the inversion pulse also dephased any transverse magnetization which it may have created. Inversion times ($\text{TI}$) were varied from 2 to 402 ms. The time between the 90$^\circ$ pulse and readout was kept constant, and therefore $\text{TR}$ ranged from 424 to 824 ms. Signal $S$ was obtained by whole-volume averaging and normalized to the maximum, $S_0$ (which corresponds to $\text{TI}\gg T_1$). $S/S_0$ was then fit to the equation

\begin{equation*}
    \ln{\left(\frac{1-S/S_0}{2}\right)}=-\frac{1}{T_1}\text{TI}
\end{equation*}

which assumes perfect inversion.

$T_2$ measurements were made using a 1D spin-echo sequence. Echo times ($\text{TE}$) were varied from 12 to 212 ms and $\text{TR}$ ranged from 1.4 to 1.6 s. Signal $S$ was obtained and normalized to the maximum $S_0$ (corresponding to $\text{TE}\approx 0$) and fit to the equation

\begin{equation*}
    \ln{(S/S_0)}=-\frac{1}{T_2}\text{TE}
\end{equation*}

assuming perfect refocusing. Crusher gradients surrounding the refocusing pulse dephased any undesired excitation. Both fits were done in Python using SciPy's \texttt{curve\_fit()} function. This function also yielded covariance estimates which were used to find uncertainties in relaxation parameters.

The $T_1$ of white mineral oil measured on our 47.5 mT MRI was $(114.153\pm 1.978)$ ms. The measured $T_2$ was $(105.087\pm 2.586)$ ms. Corresponding $R^2$ values for the $T_1$ and $T_2$ fits were calculated manually and were both 0.98.

\subsection*{S2. Supplemental Video S1}

Annotated video recording of the phantom in operation, showing the top liquid surface within the phantom housing with the manually annotated fluid level overlaid, and the CRICKIT/CPX LEDs in frame for TTL timing reference. Submitted as a separate supplementary file.

\end{document}